\documentclass{conm-p-l}

\newcommand{\calC}{{\mathcal C}}
\newcommand{\calH}{{\mathcal H}}
\newcommand{\calU}{{\mathcal U}}
\newcommand{\calL}{{\mathcal L}}
\newcommand{\bE}{{\bf E}}
\newcommand{\bP}{{\bf P}}

\newtheorem{theorem}{Theorem}[section]

\newtheorem{remark}[theorem]{Remark}

\newcommand{\nn}{\nonumber} 
\newcommand{\real}{{\bf R}}

\def\bd{\begin{displaymath}}
\def\ed{\end{displaymath}}

\def\eqref#1{(\ref{#1})} 

\def\to{\rightarrow}

\begin{document}

\title{Statistical Mechanics of Anharmonic Lattices} 
\author{Luc Rey-Bellet}
\address{Department of Mathematics, University of Virginia, 
Kerchof Hall, Charlottes\-ville, VA 22903}
\curraddr{Department of Mathematics and Statistics, University of Massachusetts, 
Lederle Graduate Research Tower, Amherst, MA 01003} 
\email{lr7q@math.umass.edu}
\urladdr{http://www.math.umass.edu/\textasciitilde lr7q}
\dedicatory{Dedicated to Lawrence E. Thomas on the occasion of his 60th birthday}
\begin{abstract} We discuss various aspects of a series of recent works on the 
non\-equi\-librium stationary states of anharmonic crystals coupled to
heat reservoirs (see also \cite{Ec}). We expose some of the main ideas and 
techniques and also present some open problems. 
\end{abstract} 
\maketitle

\section{Introduction} As emphasized in {\cite{Ru3} the study of 
nonequilibrium stationary states, i.e. the states of systems
maintained far from equilibrium by suitable forces and/or reservoirs
has seen some progress in the last few years. Unlike in equilibrium
statistical mechanics, non\-equi\-li\-brium sta\-tio\-nary states are
not given  by some apriori formula, and therefore the construction of the
stationary states and the study of their properties require in general
a thorough understanding of the dynamics. Furthermore exactly solvable
models (for example linear ones \cite{RLL,SL}) have pathological
transport properties, and this makes the study of nonlinear dynamics
even more necessary.  There are several ways to model reservoirs: by
{\em thermostats} modeled by deterministic forces, see e.g
\cite{GC,Ru1}, by {\em stochastic reservoirs} modeled by suitable
random forces, see e.g. \cite{BL}, and by {\em Hamiltonian reservoirs}
where the reservoirs are Hamiltonian systems. The last choice is
perhaps the most natural and fundamental one, but each of the
approaches involves some idealization and all of them should be in
some (yet unknown) sense equivalent. 

We consider here a class of systems (finite lattice of anharmonic
oscillators) interacting, at the boundaries only, with Hamiltonian
reservoirs described by free phonon fields. A series of rigorous
results has been obtained for such systems \cite{EPR1,EPR2, EH1, RT1,
RT2,RT3,HN,EH2}: existence and uniqueness of the stationary state,
exponential rate of convergence, positivity of entropy production and
study of its fluctuations (Gallavotti-Cohen Theorem). These results
are the most complete ones obtained up to date for a (boundary driven)
Hamiltonian system with a nontrivial (i.e. nonlinear) dynamics (see
also e.g. \cite{FGS,GKI,CELS} for classical systems and \cite{Ru2,JP}
quantum systems).  But several basic and fundamental problems remain
very poorly understood both at the physical and mathematical levels,
such as deriving the transport properties of such systems, for example
the Fourier's Law of heat conduction (see e.g. the reviews
\cite{BLR,LLP} and references therein).

In Section \ref{coupling}, we describe a model of a Hamiltonian
reservoir at positive temperature. We also describe how to choose a
coupling with the system in such a way that one can reduce the
infinite-dimensional Hamiltonian dynamics to a Markovian dynamics on a
finite-dimensional phase space.

In Section \ref{chain}, we consider a chain of oscillators connected
to two reservoirs at different temperatures and present our results on
the ergodic properties of such systems. We present some conceptual
ideas behind the proofs: the analysis of dissipation and fluctuations,
the construction of Liapunov functions for the dynamics , and the role
of breathers as (possible) obstacles to the existence and/or
exponential relaxation of the stationary state. 

In Section \ref{flow} we study the properties of the entropy
production:  positivity,  fluctuations (Gallavotti-Cohen
theorem \cite{ECM,GC,Kur,LS,Ma}), connection with  time-reversal, Green-Kubo
formula.

In Section \ref{langevin} consider stochastic reservoirs modeled by
Langevin equations. Our techniques apply to these systems too and 
we consider higher-dimensional lattice for which similar results 
as in Sections \ref{chain} and \ref{flow} can be sometimes proved. 

We would like to thank Jean-Pierre Eckmann, Martin Hairer, 
Joel Le\-bo\-witz, Claude-Alain Pillet, and Lawrence E. Thomas for their 
collaboration, suggestions, and discussions.

\section{Markovian Heat Reservoirs}\label{coupling} 
It is customary to model the interaction of a mechanical system
reservoirs by adding suitable random forces (stochastic reservoirs).
and it is often done in such a way as to obtain a Markovian dynamics.
In general the dynamics of a mechanical system interacting with
a Hamiltonian reservoir is not Markovian: there are always memory effects.  
In certain cases one can obtain a Markovian dynamics by taking a suitable
limit (see e.g. \cite{FKM,Da,Sp}), but in this section, we show how to
obtain a Markovian dynamics by choosing a {\em particular coupling}
with the reservoir. The dynamics of the system is not Markovian, but
one can make it Markovian by adding finitely many auxiliary variables.
Equations similar to those we derive do appear in \cite{Tr} but, to
our best knowledge, their derivation appear first in
\cite{EPR1,RT2}. For simplicity we consider a single particle coupled
to one reservoir.

The reservoir is a free phonon gas described by a linear wave equation
in $\real^d$. Let $\phi(x)=(\varphi(x),\pi(x))$, $x\in \real^d$, be a
pair of real fields, let $\|\phi\|$ be the norm given by $\|\phi\|^2
\,\equiv \, \int dx\, (|\nabla \varphi(x)|^2 + |\pi(x)|^2)$, and let
us denote $\langle \cdot , \cdot \rangle$ the corresponding scalar
product.  The phase space of the reservoirs at finite energy is the
real Hilbert space of functions $\phi(x)$ such that the energy
$H_B(\phi)= \|\phi\|^2/2$ is finite and the equations of motion are
\begin{equation} 
{\dot \phi}(t,x) \,=\, {\calL} \phi(t,x)\,, \quad {\calL} \,=\, \left(
\begin{array}{cc} 0 & 1 \\ \Delta & 0 \end{array}\right)\,.  
\end{equation} 
Reservoirs at positive temperature $T$ are described by Gibbs
measure at temperature $T$, $\nu_T$, given (formally) by
\begin{equation} 
\nu_T(d\varphi,d\pi) = Z^{-1} \exp{\left( -\frac{1}{2T} \int dx\, 
(|\nabla \varphi(x)|^2 + |\pi(x)|^2) \right)} \prod_x d\phi(x) d\pi(x)\,.
\end{equation}
This expression is formal, but the measure $\nu_T$ is simply the
product of a Wiener measure times a white noise measure. Its
covariance is $T\langle\cdot, \cdot\rangle$. We will construct the stationary 
states by assuming that, at time $t=0$, the reservoirs initial conditions 
are distributed according to the Gibbs measure $\nu_T$.

The Hamiltonian of the particle is $H_S(p,q) = p^2/2 + V(q)$, where
$(p,q) \in \real^d\times\real^d$, and as the Hamiltonian for the coupled
system particle and reservoir we take (dipole approximation)
\begin{eqnarray}
H(\phi,p,q) \,&=&\, \frac{1}{2}\|\phi\|^2 + p^2 + V(q) + q \cdot \int dx 
\, 
\nabla \varphi(x) \rho(x)  \\ 
\,&=&\, H_B(\phi) + H_S(p,q) + q \cdot \langle\phi , \alpha \rangle \,, \nn
\end{eqnarray} 
where $\rho(x)$ is a real rotation invariant function and 
$\alpha= (\alpha^{(1)}, \cdots, \alpha^{(d)})$ is, in Fourier space, given 
by ${\hat \alpha}^{(i)}(k) = \left( 0, -ik^{(i)} {\hat \rho}(k) / k^2 
\right)$.
Let us introduce the covariance matrix $C^{(ij)}(t) = \langle
\exp{(\calL t)}\alpha^{(i)}\,,\, \alpha^{(j)}\rangle$. A simple computation
shows that
\begin{equation}
C^{(ij)}(t)\,=\, \frac{1}{d}\delta_{ij} \int dk \, |\rho(k)|^2
e^{i|k|(t-s)}\,,
\end{equation} 
and we define a coupling constant $\lambda$ by putting 
$\lambda^2 = C^{(ii)}(0)= \frac{1}{d}\int dk  |\rho(k)|^2 $.
The equations of motion of the coupled system are
\begin{eqnarray}
{\dot q}(t) \,&=&\, p(t) \,, \nn \\
{\dot p}(t) \,&=&\, - \nabla V(q(t)) - \langle\phi, \alpha \rangle\,, \label{pl}\\
{\dot \phi}(t,k)\,&=&\, \calL \left( \phi(t,k) + q(t)
\cdot \alpha(k) \right)\,. \nn
\end{eqnarray}
With the change of variables 
$\psi(k) = \phi(k) + q \cdot \alpha(k)$, Eqs. \eqref{pl} become 
\begin{eqnarray} 
{\dot q}(t) \,&=&\, p(t)\,,  \nn\\
{\dot p}(t) \,&=&\, - \nabla V_{\rm eff}(q(t)) - \langle \psi, \alpha 
\rangle 
\label{e3}\,,  \\
{\dot \psi}(t,k)\,&=&\, \calL  \psi(t,k) 
+ p(t) \cdot \alpha(k)\,, \nn
\end{eqnarray}
where
$V_{\rm eff}(q) = V(q) -\lambda^2q^2/2$.
Integrating the last of Eqs. \eqref{e3} with initial condition $\psi_0(k)$ 
one finds
\begin{equation}
\psi(t,k)\,=\, e^{\calL t}\psi_0(k) + \int_0^tds\, e^{\calL (t-s)} 
\alpha(k)\cdot p(s)\,,
\end{equation}
and inserting into the second of Eqs. \eqref{e3} gives
\begin{eqnarray}
{\dot q}(t) \,&=&\, p(t)\,, \label{e4} \nn \\
{\dot p}(t) \,&=&\, - \nabla V_{\rm eff}(q(t)) - \int_0^t\,ds C(t-s) p(s) 
- 
\langle \psi_0, e^{-\calL t}  \alpha\rangle \label{e5}\,. 
\end{eqnarray}
Our assumption on the initial condition of the reservoirs imply that the 
force $\xi(t)=\langle\psi_0, e^{-\calL t} \alpha\rangle$ is a
$d$-dimensional stationary Gaussian process with mean $0$ and
covariance $T C(t-s)$. Note that the covariance itself appears in the
deterministic memory term on the r.h.s. of
Eq.\eqref{e5} (fluctuation-dissipation relation).

We choose the coupling function $\rho$ such that
\begin{equation}
|k|^{d-1}|\rho(k)|^2 = P(k^2)^{-1}\,,
\end{equation}
where $P$ is a polynomial. As a consequence there is a polynomial
$p(u)$ which is a real function of $iu$ and  has its roots in the lower half plane 
such that 
\begin{equation}
C^{(ii)}(t)\,=\, \int_{-\infty}^{\infty} du\, \frac{1}{|p(u)|^2}
e^{iut}\,.  
\end{equation} 
Note that this is a Markovian assumption \cite{DM}: for such couplings 
$\xi(t)$ is a Markovian Gaussian process: $p(-id/dt)\xi(t)={\dot \omega}(t)$, 
where ${\dot \omega}(t)$ is a white noise.
For simplicity we will take  $P(k^2) = C (k^2 + \gamma^2)$ and then   
$\xi(t)$ is an Ornstein-Uehlenbeck process; other polynomials can be treated similarly.  
This assumption together with the fluctuation-dissipation relation permits, by
extending the phase space with one auxiliary variable, to
rewrite the integro-differential equations \eqref{e5} as a Markov
process. We have  $C^{(ii)}(t) = \lambda^2 e^{-\gamma|t|}$ and  
introducing the variable $r$
defined by
\begin{equation}
\lambda r(t) \,=\,\int_0^t ds\, C(t-s) p(s) + \xi(t) \,,
\end{equation}
we obtain from Eqs.\eqref{e5} the set of Markovian differential
equations:
\begin{eqnarray}
{\dot q}(t) \,&=&\, p(t)\,,  \nn \\
{\dot p}(t) \,&=&\, - \nabla V_{\rm eff}(q(t)) - \lambda r(t) 
\,,\label{e8} 
 \\
{ dr}(t) \,&=&\, (- \gamma r(t) + \lambda p(t))\,dt + 
(2T \gamma)^{1/2} d\omega (t) 
\nn \,.
\end{eqnarray}
The dynamics of $(p(t),q(t),r(t))$ is Markovian, and similar equations
may be obtained for any polynomial $P$.

\section{Ergodic properties: the chain}\label{chain} 
Let us consider a chain of $n$ anharmonic oscillators given by the Hamiltonian
\begin{eqnarray}
H_S(p,q)\,=\, \sum_{i=1}^n \frac{p_i^2}{2} + V(q_1, \cdots, q_n)\,, \nn \\
V(q)\,=\, \sum_{i=1}^n U^{(1)}(q_i) + \sum_{i=1}^{n-1}
U^{(2)}(q_i-q_{i+1}) \,. \nn
\end{eqnarray}
Our assumptions on the potential $V(q)$ are

\vspace{0.5cm}
\noindent
{\bf H1 Growth at infinity}: The potentials $U^{(1)}(x)$ and
$U^{(2)}(x)$ are $\calC^\infty$ and grow at infinity like
$\|x\|^{k_1}$ and $\|x\|^{k_2}$: There exist constants $C_i$, $D_i$,
$i=1,2$ such that
\begin{eqnarray}
\lim_{\lambda \rightarrow \infty}  \lambda^{-k_i}    U^{(i)}( \lambda x ) 
\,&=&\, a^{(i)} \|x\|^{k_i} 
\,, \label{g1} \\
\lim_{\lambda \rightarrow \infty}  \lambda^{-k_i +1} \nabla U^{(i)}( 
\lambda x ) 
\,&=&\, a^{(i)} k_i 
\|x\|^{k_i-2}  x \,, \label{g2} \\
\| \partial^2 U^{(i)}(x) \| \, &\le& \, 
( C_i + D_i V(x))^{1-\frac{2}{k_i}}  \,. \label{xcx} 
\end{eqnarray}
where $\| \cdot\|$ in Eq. \eqref{xcx} denotes some matrix-norm. 

Moreover we will assume that
\begin{equation}
k_2 \,\ge \, k_1 \, \ge \, 2\,,
\end{equation}
so that, for large $\|x\|$ the interaction potential $U^{(2)}$ is
"stiffer" than the one-body potential $U^{(1)}$. 

\vspace{0.5cm}
\noindent
{\bf H2 Non-degeneracy}: The coupling potential between nearest
neighbors $U^{(2)}$ is non-degenerate in the following sense.  For
$x\in {\bf R}^d$ and $m=1,2, \cdots$, let $A^{(m)}(x): {\bf R}^d
\rightarrow {\bf R}^{d^{m}}$ denote the linear maps given by
\begin{equation}
(A^{(m)}(x) v)_{l_1 l_2 \cdots l_{m}} \,=\, 
\sum_{l=1}^d \frac{\partial^{m+1}U^{(2)}}{\partial x^{(l_1)} 
  \cdots \partial x^{(l_m)} \partial x^{(l)}}(x) v_l \,.
\end{equation}
We assume that for each $x \in {\bf R}^d$ there exists $m_0$ such that 
\begin{equation}
{\rm Rank} ( A^{(1)}(x), \cdots A^{(m_0)}(x)) = d \,.
\end{equation}

For example any confining polynomial potential of even degree satisfy 
assumptions {\bf H1} 
and {\bf H2}.

We couple the first and $n^{th}$ particle each to one reservoir at 
temperatures $T_1$ and $T_n$ respectively. We assume that the couplings 
to be as in Section \ref{coupling} so that, by introducing two auxiliary 
variables $r_1$ and $r_n$, we obtain the set of stochastic differential 
equations equations
\begin{eqnarray}
{\dot q_1} \,&=&\, p_1\,, \label{e15} \nonumber \\
{\dot p_1} \,&=&\, - \nabla_{q_1} V(q) - \lambda r_1 
\,,\label{e16} 
\nonumber \\
{ dr_1} \,&=&\, (- \gamma r_1 + \lambda p_1)\, dt + 
(2T_1 \gamma)^{1/2} d\omega_1\,, \label{e17} \nonumber \\
{\dot q_j} \,&=&\, p_j\,, \quad\quad \quad \quad\quad j=2,\dots,n-1\,, 
\label{e18} \nonumber \\
{\dot p_j} \,&=&\, - \nabla_{q_j} V(q)\,, \,\,\quad j=2,\dots,n-1 \,, 
\label{e22}  \\
{\dot q_n} \,&=&\, p_n \,, \label{e20}\nonumber  \\
{\dot p_n} \,&=&\, - \nabla_{q_n} V(q) - \lambda r_n 
\,,\label{e21} 
\nonumber \\
{ dr_n} \,&=&\, (- \gamma r_n + \lambda p_n)\,dt + 
(2T_n \gamma)^{1/2} d\omega_n \,.\nn
\end{eqnarray}
The solution $x(t)=(p(t),q(t),r(t))$ of Eq.\eqref{e22} is a Markov process.  
We denote $T^t$ as the associated semigroup, 
\begin{equation} 
T^tf(x) \,=\, \bE_x [ f(x(t)]\,, 
\end{equation} 
with generator 
\begin{eqnarray}
L\,&=&\, \sum_{i \in \{1,n\}}  
\gamma \left( \nabla_{r_i} T \nabla_{r_i} - r_{i} \nabla_{r_i}\right) + 
\lambda \left( p_i \nabla_{r_i} - r_i \nabla_{p_i} \right)  \\
&&+ \sum_{i=1}^n p_i \nabla_{q_i} - (\nabla_{q_i} V(q)) \nabla_{p_i} \,,\nn
\end{eqnarray} 
and $P_t(x,dy)$ as the transition probability of the Markov
process $x(t)$.  
There is a natural energy function which is associated to
Eq.\eqref{e22}, given by 
\begin{equation}\label{g} 
G(p,q,r)\,=\, \frac{r^2}{2} + H(p,q)\,,  
\end{equation} 
and a
straightforward computation shows that at equilibrium (i.e., if $T_1=T_n=T$) 
the Gibbs measure
$Z^{-1} \exp{(- G(p,q,r)/T)}$ 
is an invariant measure for the Markov process $x(t)$.  

The function 
\begin{equation}\label{wtheta}
W_\theta \,=\, \exp{(\theta G)}
\end{equation}
will be used repeatedly. 
We denote as $| \cdot |_\theta$ the weighted total variation norm 
given by 
\begin{equation}\label{wvar} 
| \pi |_\theta \,=\, \sup_{|f| \le W_\theta} \left| \int
f d\pi \right |\,, 
\end{equation}
for any (signed) measure $\pi$.  
We introduce norms $\|\cdot\|_\theta$ and
Banach spaces $\calH_\theta$ given by 
\begin{equation}\label{ltheta}
\|f\|_\theta \,=\, \sup_{x\in X} \frac{|f(x)|}{W_\theta(x)}\,, \quad
\calH_\theta\,=\, \{ f \,:\, \|f\|_\theta < \infty \}\,,
\end{equation} 
and write $\|K\|_\theta$ for the norm of an operator $K:
\calH_\theta \rightarrow \calH_\theta$. 

Our results on the ergodic properties of Eqs. \eqref{e22} are summarized in 
\begin{theorem}{\bf : Ergodic properties} \label{main} {\hfill} 

\vspace{0.2cm}
\noindent
(a) The Markov process $x(t)$ has a unique invariant measure $\mu$. The 
measure $\mu$ is ergodic and mixing and has a $\calC^\infty$ everywhere 
positive density. 

\vspace{0.2cm} 
\noindent
(b) For any $\theta$ with $0 < \theta < (\max\{T_1,T_n\})^{-1}$ the 
semigroup $T^t:\calH_\theta \rightarrow \calH_\theta$ is compact for all 
$t>0$. In particular the process $x(t)$ converges exponentially 
fast to its stationary state $\mu$: there exist constants 
$r=r(\theta) > 1$ and $R=R(\theta) <\infty$  such 
that 
\begin{equation} \label{s01}
 |P_t(x,\cdot) - \mu|_\theta  \, \le \, R r^{-t} W_\theta(x) \,,
\end{equation}
for all $x \in X$.  
Furthermore for all functions $f$, $g$ 
with $f^2$, $g^2 \in \calH_\theta$ and all $t>0$ we have 
\begin{equation} 
\left| \int g T^tf \,d\mu  - \int f \,d\mu \int g \,d\mu \right|  \,\le 
\, R r^{-t} \|f^2\|_\theta^{1/2}
\|g^2\|_\theta^{1/2}\,,
\end{equation}
(exponential decay of correlations in the stationary state). 
\end{theorem}

There are essentially two proofs of Theorem \ref{main} which 
follow quite different strategies.  The (chronologically) first one 
\cite{EPR2,RT2,RT3} is probabilistic and is based on a detailed analysis of 
the dynamical effects of dissipation and fluctuations. The second proof 
\cite{EPR1,EH1,HN,EH2} is purely functional-analytic and use global 
hypoelliptic estimates on the generator $L$. In our opinion the first approach 
has the advantage to be a little simpler and also to display better 
the dynamical mechanisms at play. It singles out the breathers as a (possible) 
obstacle to the existence of a stationary state. The second approach has 
the advantage to be a bit more constructive and so it gives more 
precise information on the location of the spectrum. 
We sketch the main steps of the proof of Theorem \ref{main}, following the 
probabilistic approach taken in \cite{EPR2,RT2,RT3}. 

\vspace{0.2cm}
\noindent {\bf Hypoellipticity:} 
The first thing to realize is that the generator $L$ is hypoelliptic. The 
operator $L$ has the form $L = \sum_{j=1}^M X_j^2 + X_0$ where the $X_i$ are 
smooth vector fields which satisfy the following H\"ormander-type 
condition: the vector fields $\{X_j\}_{j\ge 1}; \{ [X_i, X_j] \}_{i,j \ge 0}; 
\{ [[X_i, X_j] X_k] \}_{i,j,k \ge 0}; \cdots $ span the whole tangent 
space at every point $x$. This implies \cite{Ho,IK,No} that the the transition 
probabilities have a smooth density. 

\vspace{0.2cm}
\noindent{\bf Control theory and Uniqueness:} 
The {\em uniqueness} of the invariant measure is obtained with a 
control-theoretic argument. Using the Support Theorem of \cite{SV}, one 
study the following control problem: replace in  Eqs. \eqref{e22} the 
(rough) white noise ${\dot \omega}(t)$ by a smoother (e.g. piecewise smooth) 
control. The Support Theorem asserts in particular that the support of the 
transition probabilities $P_t(x,dy)$ is the same as the closure of the set of all points 
reachable in a time $t$ starting from $x$ with a smooth control.  One can show 
\cite{EPR2} that for all $t$ and for all $x$ the transition probabilities 
have full support. This is achieved done by an explicit construction of the 
possible controls which drive the system from $x$ to $y$ in a given time span.
This controllability property together with the smoothness of the transition 
probability imply that there is at most an invariant measure. 

\vspace{0.2cm}
\noindent{\bf Liapunov functions, Existence, and Compactness:}
The {\em existence} of the invariant measure is the most
difficult property to establish. This is in sharp contrast with
equilibrium where the invariant measure is given, a priori, by the
usual Gibbs Ansatz. Also as we prove the existence of an invariant measure
we prove strong ergodic properties which, of course, are also of
interest at equilibrium.  The idea is to construct a Liapunov function
for the Markov process $x(t)$ (for a detailed exposition of the
subject see \cite{MT}). For this problem we proved the following.
Fix $t>0$ and $\theta < (\max\{T_1,T_n\})^{-1}$ and consider the
function $W_\theta(x)$ given by Eq. \eqref{wtheta}. There exist a
constant $E_0$ and  functions $\kappa(E)$ and $b(E)$ defined on $[E_0,\infty)$ 
with $\lim_{E \to \infty} \kappa(E) =0$such that for $E> E_0$
\begin{equation}\label{liapunov}
T^t W_\theta(x) \,\le\, \kappa(E) W_\theta(x) + b(E) \chi_{\{G \le E\}}(x)\,.
\end{equation}
This means that, outside the compact set $\{G \le E\}$, the dynamics
is dissipative and since $\kappa(E)$ tends to zero as $E$ tends to
infinity, the dissipation at high energies can be made arbitrarily
strong. This particular property of the Liapunov function together with 
the smoothness of the transition probabilities implies that
the semigroup $T^t$ is compact on the Banach spaces $\calH_\theta$.

One should think of the dynamics as follows: to the conservative
Hamiltonian dynamics two forces are added, a dissipative force (the
terms $-\gamma r_i$ in \eqref{e22}) and a random force (the white
noises) which are proportional to the temperature. At high energies,
i.e., at energies much bigger than the temperatures of the reservoirs,
dissipation dominates and there is a strong drift which let the energy
of the system decrease and the fluctuating forces are negligible
compared to the dissipation.  At low energies, on the contrary the
fluctuations are not dominated by the dissipation anymore. 

Since both dissipation and noise act only at the boundary of the
chain, a key ingredient in the analysis is a bound on propagation of energy 
in anharmonic lattices. Think for example of an initial condition in
which the energy is concentrated in one oscillator far away from the
boundary. To prove Eq. \eqref{liapunov} one must have a lower bound on
how much of the energy propagates through the chain to the boundary to
get dissipated. For this bound the condition {\bf H2} is crucial. It
is well-known \cite{ST,MA,Ba} that in networks of anharmonic
oscillators there are breathers which are (Nekhoroshev-stable) time
periodic exponentially localized solutions of the Hamiltonian
equations of motions. A simple scaling argument shows that the high
energy behavior of breathers is very different depending on whether
the condition {\bf H2} is satisfied or not. If it is not satisfied,
the higher the energy, the more localized the breather tends to be: at
high energy we can have states with oscillators oscillating
very fast and barely interacting with their neighbors which are
essentially at rest. On the contrary, if Condition {\bf H2} is
satisfied, then one can show that for any initial condition of sufficiently large
energy $E$, the kinetic energy of the oscillator on the boundary will be at
least of order $E^{2/k_2}$ on a time interval of order $1$.

To prove \eqref{liapunov}, in a first step one sets $T_1=T_n=0$ in
\eqref{e22} and one obtains a set of deterministic ODE's equations. 
Using the bound on propagation of energy one shows that for large 
enough $G(0)=E$ we have
\begin{equation}\label{detdiss}
G(1) - G(0) \le - c E^{2/k_2}\,. 
\end{equation}
Note that this corresponds to the physical situation where the
reservoirs, at time $t=0$ are at energy $0$. In this case the system
simply radiates all its energy into the reservoirs and relax into a
state corresponding to a stationary point of the Hamiltonian $H$. If the temperatures 
are non-zero, one shows that on a suitable time interval the random solution of 
\eqref{e22} do follow closely the deterministic trajectories with very high 
probability. So for most trajectories, an estimate of the type 
\eqref{detdiss} also holds. To conclude of the proof of \eqref{liapunov}, one 
considers the function $W_\theta= \exp{(\theta G)}$ and uses some stochastic analysis 
(see \cite{RT2} for details).

\section{Heat Flow and Entropy Production}\label{flow}
To define the heat flow and the entropy production we write the energy of the 
chain $H$ as a sum of local energies $H= \sum_{i=1}^n H_i$ where
\begin{eqnarray} 
H_1\,&=&\, \frac{p_1^2}{2} + U^{(1)}(q_1) + \frac{1}{2} 
U^{(2)}(q_{1}-q_1) \,, \nn \\
H_i \,&=&\, \frac{p_i^2}{2} + U^{(1)}(q_i) + 
\frac{1}{2} \left(U^{(2)}(q_{i-1}-q_i) + U^{(2)}(q_{i}-q_{i+1})\right) \,,  
\label{hi} \\
H_n\,&=&\, \frac{p_n^2}{2} + U^{(1)}(q_n) + 
\frac{1}{2} U^{(2)}(q_{n}-q_{n-1}) \,. \nn 
\end{eqnarray} 
Differentiating with respect to time one finds
\begin{equation}
\frac{d}{dt}T^t H_i\,=\, T^t (\Phi_{i-1} - \Phi_i)\,,
\end{equation}
where 
\begin{eqnarray} 
\Phi_0\,&=&\,  -\lambda r_1p_1 \,,\nn \\
\Phi_i \,&=&\, \frac{(p_{i} + p_{i+1})}{2} \nabla U^{(2)}(q_{i}-q_{i+1}) \,,
\label{fii} \\
\Phi_n\,&=&\,  \lambda r_np_n  \,. \nn 
\end{eqnarray} 
It is natural interpret $\Phi_i$, $i=1,\cdots,n-1$ as the heat flow from the 
$i^{th}$ to the $(i+1)^{th}$ particle, $\Phi_0$ as the flow from the left 
reservoir into the chain, and $\Phi_n$ as the flow from the chain into the 
right reservoir. We define corresponding entropy productions by 
\begin{equation} 
\sigma_i \,=\, \left(\frac{1}{T_1} - \frac{1}{T_n}\right) \Phi_i \,. 
\end{equation} 
There are other possible definitions of heat flows and corresponding
entropy production that one might want to consider.  One might, for
example, consider the flows at the boundary of the chains, and define
$\sigma_b=\Phi_1/T_1 - \Phi_n/T_n$. Also our choice of local energy is
somewhat arbitrary, other choices are possible but this does not
change the subsequent analysis.

Our results on the heat flow are summarized in
\begin{theorem}{\bf : Entropy production}\label{entprod}

\vspace{0.2cm} 
\noindent
(a) The expectation of the entropy production $\sigma_j$ in the stationary
state is independent of $j$ and nonnegative
\begin{equation}\label{nonneg}
\int \sigma_j d\mu \, \ge\, 0\,,
\end{equation}
and it is positive away from equilibrium
\begin{equation}\label{pos}
\int \sigma_j d\mu \, = \, 0 \qquad {\rm ~if~and~only~if~} \qquad T_1=T_n \,.
\end{equation}

\vspace{0.2cm}
\noindent
(b) The ergodic averages 
\begin{equation}
\frac{1}{t}\int_0^t \sigma_j (x(s)) 
\end{equation}
satisfy the large deviation principle: There exist a neighborhood $O$ of the 
interval $[-\int \sigma_j d\mu, \int \sigma_j d\mu]$ and a rate function 
$e(w)$ (both independent of j) such that for all intervals $[a,b] \subset O$ we have 
\begin{equation}
\lim_{t \to \infty} - \frac{1}{t}
\log \bP_x\{{\overline \sigma_i}^t \in [a,b]\} \,= \, \inf_{w \in [a,b]} 
e(w)\,.
\end{equation}
Moreover the rate function $e(w)$ satisfy the relation
\begin{equation}\label{gc}
e(w)-e(-w) \,=\, - w \,,
\end{equation}
i.e., the odd part of $e$ is linear with slope $-1/2$. 
\end{theorem}
Let us consider the functions $R_j$ given by
\begin{equation} \label{ri}
R_{j}\,=\, \frac{1}{T_1} \left(\frac{r_1^2}{2} + \sum_{k=1}^{j}
H_k(p,q)\right) + \frac{1}{T_n} \left( \sum_{k=j+1}^{n} H_k(p,q) +
\frac{r_n^2}{2} \right) \,,  
\end{equation} 
so that $\exp{(-R_j)}$ is a ``two-temperatures'' Gibbs state. We
denote by $J$ the time reversal operator which changes the sign of the
momenta of all particles $Jf(p,q,r)=f(-p,q,r)$ and we denote as $L^T$ the 
formal adjoint of the generator $L$ (the Fokker-Planck operator).
The following (formal) operator identities are easily verified
\begin{equation}\label{conj0} e^{R_j} J L^T J e^{-R_j}\,=\, L - \sigma_j \,, 
\end{equation} 
and also for any constant $\alpha$
\begin{equation}\label{conj} e^{-R_j} J ( L^T - \alpha \sigma_j) J
e^{R_j}\,=\, L - (1-\alpha) \sigma_j \,.  
\end{equation} 
These identities are the key element to prove both Eqs. \eqref{nonneg}
and \eqref{gc}. The fact that the entropy production is strictly positive 
away from equilibrium do require more work \cite{EPR2}. 

Let us sketch the proof of Eqs. \eqref{nonneg}. We write the 
positive density $\rho(x)$ of $\mu(dx)=\rho(x)dx$ as 
\begin{equation}
\rho  = J e^{-R_j} e^{-F_j}\,. 
\end{equation}
Let $L^*$ denote the adjoint of $L$ on $L^2(\mu)$, it is given by 
$L^* = \rho^{-1} L \rho$ and using Eq. \eqref{conj0} a simple computation 
shows that
\begin{eqnarray}
JL^*J \,&=&\, e^{F_j} (L - \sigma_j)  e^{-F_j} \nn  \\
\,&=&\, L - \sigma_j - (LF_j) - 2 \sum_{i \in \{1,n\}} T_i (\nabla_{r_i}F_j) 
\nabla_{r_i} \label{gh} \\ 
&&\, + \sum_{i \in \{1,n\}} ( T_i |\nabla_{r_i}F_j|^2) \,. \nn
\end{eqnarray}
The operator $JL^*J$ is the generator of the time-reversed process and 
using that $L^*1=0$ we find the identity
\begin{equation}
\sigma_i\,=\, \sum_{i \in \{1,n\}} T_i |\nabla_{r_i}F_j|^2  - LF_i\,. 
\end{equation} 
The first term is obviously positive while the expectation of the second 
term in the stationary state vanishes and so we obtain Eq. \eqref{nonneg}.

Let us turn now to Eq. \eqref{gc}. Let us give first a formal proof, 
ignoring any technicality (the argument is essentially from \cite{Kur}). 
To the study of the large deviations the large deviations of 
$t^{-1}\int_0^t \sigma_i (x(s)) ds$ we consider the moment generating 
functionals 
\begin{equation} 
\Gamma^i_x(t,\alpha) \,=\, \bE_x \left[ e^{- \alpha
\int_0^t \sigma_i (x(s))\, ds } \right] \,.  
\end{equation} 
Formally the Feynman-Kac formula gives 
$\Gamma^i_x(t,\alpha)=e^{t(L-\alpha\sigma_i)}1(x)$ and the large deviation 
functional $e(w)$ is given by the Legendre transform of the function
\begin{equation}
e(\alpha) \equiv \lim_{t \to \infty} -\frac{1}{t} \log 
\Gamma^i_x(t,\alpha) \,.
\end{equation}
By a Perron-Froebenius argument $e(\alpha)$ is the largest eigenvalue 
of $L-\alpha \sigma_i$. Since $L-\alpha \sigma_i$ is conjugated to 
$L^T-(1-\alpha) \sigma_i$ by Eq. \eqref{conj} and since $L-\alpha \sigma_i$ and 
$L^T-\alpha \sigma_i$ should have the same spectrum we conclude that
$e(\alpha) = e(1-\alpha)$. 
Taking Legendre transform we obtain Eq. \eqref{gc}.

How do we make this argument rigorous? From the form of  the entropy 
production $\sigma$ (it is an unbounded function), one sees that $L- \alpha \sigma_i$ is 
not a relatively bounded perturbation of $L$.  
A priori it is not even obvious $L- \alpha \sigma_i$ is the generator 
of a semigroup, i.e., that the function 
$\Gamma^i_x(t,\alpha)$ is finite for $\alpha\not= 0$.  
To make things work we will use the following identity which can be checked 
easily
\begin{equation}\label{op} L-\alpha \sigma_i \,=\, e^{\alpha R_i}
{\overline L}_\alpha e^{-\alpha R_i}\,, \end{equation} 
where
\begin{equation}\label{lba} 
{\overline L}_\alpha \,=\, 
L + \gamma \sum_{i \in \{1,n\}} \left( 2\alpha r_i \nabla_{r_i} 
-(\alpha - \alpha^2) T_{i}^{-1} r_i^2\right) + 2d \gamma \alpha \,.
\end{equation}
This shows that all the operators $L-\alpha\sigma_i$ are conjugated to the 
same operator ${\overline L}_\alpha$. It is not hard to see that 
${\overline L}_\alpha$ is relatively bounded perturbation of $L$. 
Using the same techniques as the 
one used in the proof of Theorem \ref{main} one can show that 
$\exp{(t {\overline L}_\alpha)}$ defines a quasibounded compact semigroup 
on $\calH_\theta$ provided that $-\alpha < \theta T_i < 1-\alpha$ and 
using Eq. \eqref{lba} this shows that the function $e(\alpha)$ exists and 
is real analytic provided 
\begin{equation}\label{i99}
\alpha \in \left( - \frac{ T_{\min} }{ T_{\max}  - T_{\min} } \,,\, 
1+  \frac{ T_{\min} }{ T_{\max} - T_{\min}}  \right)\,.
\end{equation}
Using Gaertner-Ellis Theorem concludes the proof of Theorem \ref{entprod}.

\begin{remark}{\bf Time reversal and entropy production I.} {\rm
It is instructive to give a interpretation of Eq. \eqref{gh} in terms of 
the path space measure of the process (see \cite{Ma}). At 
equilibrium ($T_1=T_n=T$) this is simply detailed balance, $R_i=G$, $JL^*J=L$ 
or $p_t(Jy,Jx) = p_t(x,y) \exp{(-(G(y)-G(x))/T)}$. If we are away from 
equilibrium we can interpret it as follows. Let $P_{st}$ denotes the 
path space measure of the stationary process starting in the state $\mu$ 
and let $\theta$ denote the operator defined by $\theta x(t)= x(-t)$. The 
path space measure of the (stationary) time reversed process is given by 
$P^{rev}_{st}=JP_{st}\circ \theta^{-1} $ and its generator is $JL^*J$. Using 
the relation \eqref{gh} and Feynman-Kac formula we find
\begin{equation}
\frac{dP_{st}^{rev}}{dP_{st}}(x(t)) \,=\, e^{-\int_0^t\, ds \sigma_j 
(x(s))} \times  e^{-(F_j(x(t)) - F_j(x(0)))}\,.
\end{equation}
As pointed out in \cite{Ma}, this gives a microscopic definition of the 
entropy production and relates it directly to the action of time-reversal. 
}
\end{remark}

\begin{remark}{\bf Time reversal and entropy production II.}
{\rm If we go back to the original Hamiltonian description of the system, 
we note that the dynamics and the in ital conditions of the reservoirs are 
invariant under time reversal: changing $t$ into $-t$, and changing
the signs of the momenta $p$ of the crystal and of the fields $\Pi$
leaves the equations of motion unchanged. Also the initial condition
of the reservoirs are distributed according to the Gibbs measure and
are invariant under reversal of the velocities in the reservoirs.
Consequently we can study the behavior of the system as $t \to -
\infty$ simply by changing $p$ into $-p$, $\pi$ into $-\pi$ and
considering $t \to \infty$. If we do this and reduce the dynamics as
in the section \ref{coupling} we obtain a Markov process with a
generator given by $JLJ$ (the variables $r$ are left unchanged) and
the system relaxes into the stationary state $J\mu$. 

Since $\sigma_j$ is an odd function of $p$, we have $\int \sigma_j d\mu 
\le 0$. One might be tempted to draw the conclusion that, in the
far distant past, heat was flowing from the cold reservoir into the hot
one. This is incorrect since the very definition of the heat flows involves 
a time-derivative, if we consider $t \to - \infty$ one should change the 
definition of the flows accordingly. 
}
\end{remark}

\begin{remark}{\bf Green-Kubo formula.}
{\rm As noted in \cite{Ga,Kur} one can derive the Green-Kubo 
from the fluctuation theorem. Here the external ``field'' is the inverse temperature difference 
$\Delta \beta = (\beta_n - \beta_1)$ and we have $\sigma_j = \Delta \beta \phi_j$. 
We consider the function $f(a, \Delta \beta)$ given by 
\begin{equation}
f(a , \Delta \beta) \equiv \lim_{t \to \infty} -\frac{1}{t} 
\log \bE_\mu \left[ e^{- a \int_0^t \phi_i (x(s))\, ds } \right] \,,  
\end{equation}
where $a=\alpha \Delta \beta$ and the second variable in $f$ indicates 
the dependence of the dynamics of the stationary state $\mu$ on $\Delta \beta$. 
From the compactness properties of the semigroups involved it is easy to see
that $f(a, \Delta \beta)$ is a real-analytic function of both variables $a$ and $\Delta
\beta$.  The relation $e(\alpha) = e(1-\alpha)$ now reads
\begin{equation} f(a,\Delta\beta) \,=\, f(\Delta \beta - a, \Delta
\beta) \,.
\end{equation}
Differentiating this relation we find
\begin{eqnarray}
\frac{\partial^2 f}{\partial a \partial (\Delta \beta)}(0,0) \,=\, 
- \frac{\partial^2 f}{\partial a \partial (\Delta \beta)}(0,0)
- \frac{\partial^2 f}{\partial a^2}(0,0) \,.
\end{eqnarray}
We have 
\begin{eqnarray} 
\frac{\partial f}{\partial a}(0,\Delta \beta) \,&=&\, \int \phi_j d\mu \,,  \\
\frac{\partial^2 f}{\partial a^2} (0,0) \,&=&\ -\frac{1}{2} \int_0^\infty \left(
\int (T_0^t \phi_j) \phi_j d\mu_O \right) \,ds \,,
\end{eqnarray}
where $T^t_0$ is the semigroup at equilibrium ($\Delta_\beta=0$) and $\mu_0$ is the 
equilibrium measure $Z^{-1} \exp{(-\beta G)}$. We obtain
\begin{equation}
\frac{\partial}{\partial (\Delta \beta)}\left.\left(\int \phi_j d\mu\right) 
\right|_{\Delta \beta=0}  \,=\,\frac{1}{2}  
\int_0^\infty \left(\int (T^t \phi_j) \phi_j d\mu \right) \,ds\,, 
\end{equation}
and this is the familiar Green-Kubo formula. 
}
\end{remark}

\section{Langevin equations and other lattice of oscillators}
\label{langevin} 

\subsection{Hypercubes}
The reader may wonder why we are only considering one-dimensional
lattice of oscillators. Although it is not difficult to construct
Hamiltonian models of higher dimensional lattice of oscillators
interacting with reservoirs, the reduction to a tractable set of SDE's 
is not trivial. In particular it is not easy to prove that reservoir 
provide enough noise/dissipation to carry the analysis done to prove 
Theorem \ref{main}.

If we consider hypercubes of oscillators and stochastic reservoirs given by Langevin 
equations much can be said. For $i \in {\bf Z}^d$, let 
$|i|=\sup_{1\le k} \le |i_k|$ and let  $\Lambda$ 
be the hypercube $\Lambda = \{ i \in {\bf Z}^d\;,\; |i| \le N \}$. The side 
of the cube $N$ is arbitrary, but finite.
On each site of $\Lambda$ there is an oscillator with coordinates $(p_i,q_i) 
\in {\bf R}^d \times {\bf R}^d$ and the Hamiltonian of the system is given 
by
\begin{equation}
H(p,q) \,=\, \sum_{i \in \Lambda} \left( \frac{p_i^2}{2} + U^{(1)}(q_i) + 
\sum_{j\in \Lambda \,:\; |j-i|=1} U^{(2)}(q_i-q_j) \right)\,.
\end{equation}
As a model of reservoirs add an Ornstein-Uehlenbeck process to each oscillator on two 
opposite sides of the hypercube $\Lambda$. The equations of motions are
\begin{eqnarray}
{\dot q_i} \,&=&\, p_i \,, \nn \\
 dp_i \,&=&\, \left(  -\nabla_{q_i} V(q) - \lambda p_i \right) dt 
+ \sqrt{ 2 \lambda T_L} d\omega_i\,, \quad {\rm ~if~} i_{1} = -N\,, \label{l}\\
{\dot q_i} \,&=&\, p_i \,, \nn \\
dp_i \,&=&\, \left(  -\nabla_{q_i} V(q) - \lambda p_i \right) dt 
+ \sqrt{ 2 \lambda T_R} d\omega_i\,, \quad {\rm ~if~} i_{1} = N\,, \label{r}\\ 
{\dot q_i} \,&=&\, p_i \,, \nn \\
{\dot p_i} \,&=&\,  -\nabla_{q_i} V(q)\,,   \quad {\rm ~if~} i_{1} \not= -N,N \,. 
\label{m}
\end{eqnarray}
We denote by $x(t)=(p(t),q(t))$ the Markov process which solves 
Eqs. \eqref{l}--\eqref{m}. We have, similarly to Theorem \ref{main}

\vspace{0.2cm}
\noindent{\bf Ergodic properties:}
The Markov process $x(t)$ has a unique invariant measure $\mu$. The
measure $\mu$ is ergodic and mixing and has a $\calC^\infty$
everywhere positive density. The convergence to the stationary state
occurs exponentially fast.
\vspace{0.2cm}

We can also prove a result analogous to Theorem \ref{entprod} by
considering the heat flow through an hypersurface $\{i_1 = k\}$. 
We define the energy of the oscillators in the hyperplane $\{i_1 = k\}$ to be 
\begin{equation}
H_k(p,q) \,=\, \sum_{i\,:\, i_1=k} \left( \frac{p_i^2}{2} + U^{(1)}(q_i) + 
\frac{1}{2} \sum_{j\in \Lambda \,:\;j_1=k,  |j-i|=1} U^{(2)}(q_i-q_j) \right)\,,
\end{equation} 
then the total heat 
flow $\Phi_k$  between the hypersurfaces $\{i_1 = k\}$ and $\{i_1 = k+1\}$ is given by 
\begin{equation} 
\Phi_k \,=\, \sum_{i: i_1=k}  \frac{p_i+p_{i+e_1}}{2} \nabla U^{(2)}(q_i - q_{i+e_1})\,,  
\end{equation}
where $e_1 = (1,0,\cdots,0)$. The corresponding entropy production is defined by 
$\sigma_k \,=\,(T_R^{-1}-T_L^{-1})\Phi_k$. As in Section \eqref{entprod} we have 

\vspace{0.2cm}
\noindent{\bf Positivity of the entropy production:}
The expectation of the entropy production $\sigma_k$ in the stationary
state is independent of $j$ and nonnegative
\begin{equation}
\int \sigma_k d\mu \, \ge\, 0\,,
\end{equation}
and it is positive away from equilibrium
\begin{equation}
\int \sigma_k d\mu \, = \, 0 \qquad {\rm ~if~and~only~if~} \qquad T_1=T_n \,.
\end{equation}

\vspace{0.2cm}
\noindent{\bf Gallavotti-Cohen fluctuation theorem:}
The ergodic averages 
\begin{equation}
\frac{1}{t}\int_0^t \sigma_k (x(s)) 
\end{equation}
obey the large deviation principle with a rate function $e(w)$ 
which satisfy the relation
\begin{equation}
e(w)-e(-w) \,=\, - w \,.
\end{equation}
\vspace{0.2cm}

The techniques used to prove Theorem \ref{main} can be used to analyze  
Eqs. \eqref{l}--\eqref{m}. 
To see this just think of the set of all 
oscillators on the hyperplane $i_1 = k$ as one $(2L+1)^{d-1}$-dimensional 
oscillator $Q_k$ with one-body potential 
\begin{equation}
W^{(1)}(Q_j)= \sum_{i\;:\; i_1=k}  U^{(1)}(q_i) + \sum_{i,j\;:\; i_1=j_1=k}  
U^{(2)}(q_i-q_j)\,,
\end{equation}
and two-body potential  
\begin{equation}
W^{(2)}(Q_k -Q_{k-1})= \sum_{i,j\;:\; i_1=k, j_1=k-1} 
U^{(2)}(q_i-q_j)\,.
\end{equation} 
Doing this we obtain a ``chain'' of (high-dimensional)
oscillators. The noise and dissipation are slightly different, but
this can be analyzed using exactly the same methods.

\subsection{General Graphs} A natural problem, considered in 
\cite{MNV,Za}, is the following. Consider an arbitrary graph $G=(V,E)$, 
at each vertex $i$ of $G$ there is an oscillator with coordinates
$(p_i,q_i)$ and energy $p_i^2/2 + U^{(1)}(q_i)$ and for each edge $e\in E$  
between $i$ and $j$ the two oscillators interact via a
two-body potential $U^{(2)}(q_i -q_j)$. The boundary of the graph is a
subset $\partial V$ of the set of vertices $V$ and if $i \in \partial
V$, an Ornstein Uehlenbeck modeling the interaction with the reservoir
is added to the Hamiltonian equation.

Given such a graph, one might ask under which condition on the graph
the analysis done for the chain can be carried over. This is largely
an open problem but the following can be said

\vspace{0.2cm}
\noindent{\bf Quadratic potentials:}
 If the potentials are quadratic, then the SDE's are linear and
necessary and sufficient conditions are known for the existence and
uniqueness of stationary states. It can be given in purely algebraic
terms, see e.g. \cite{IK}. 

\vspace{0.2cm}
\noindent{\bf Nonuniqueness of the stationary state:}
On the other hand, linear models have (too) many invariants and very
simple model have more than one invariant measure, e.g., four
harmonic oscillators arranged in a diamond shape with two opposite oscillators
connected to heat baths have a one-parameter family of invariant
measures \cite{Za,MNV}. 

\vspace{0.2cm}
\noindent{\bf Entropy production:}
The nonnegativity of the entropy production is very easy to
establish at the formal level. Let $R$ be the function given by $R=
\sum_{i \in \partial V} (T_i)^{-1} p_i^2/2$ where $T_i$ is the
temperature of the reservoirs attached to the oscillator $i$.  We have
the relation $e^R J L^T j e^{-R} = L- \sigma$, where $\sigma = \sum_{i
\in \partial V}T_i^{-1}(p_i^2-T_i)$. The quantity is $(T_i-P_i^2)$ is
to be interpreted as the heat flow from the system into the
corresponding reservoir at temperature $T_i$. As in section
\ref{entprod} one shows that this relation implies the nonnegativity
of the entropy production in the stationary state. Using conservation
laws other entropy productions involving heat flows through the bulk of the 
lattice can be considered too.

\vspace{0.2cm}
\noindent{\bf Positivity of the entropy production:}
In \cite{MNV} an interesting condition is given which implies
the positivity of the entropy production, provided one assumes
the existence of a smooth positive stationary state.  As mentioned
above, if the elements of a graph are connected to many others, the
potential should be ``sufficiently'' non-linear. The condition is the
following: a function $f$ is called $n$-nondegenerate provided the set
\begin{equation}
\calU^n \,=\, \left\{ (q_1,\cdots,q_n) \in {\bf R}^n :  
\exists (q'_1, \cdots, q'_n)\in {\bf R}^n : 
\det(f(q'_i -q_j)) \not=0 \right\}\,,
\end{equation}
is dense in ${\bf R}^n$. For example the polynomials of degree $r$ are
$n$-nondegenerate provided $r\ge n-1$. One can show (\cite{MNV} for details)
that if $d^2 U^(2)/dq^2$ is $n$-nondegenerate for $n$ sufficiently
large then the entropy production is non-negative. How large $n$
should be depends on the graph, in particular on how many nearest
neighbors an oscillator has and if there are many loops in the
graph. We conjecture that such a condition should also imply existence
and uniqueness of the stationary state.

\end{document}